\documentclass[10pt, final, a4paper]{iopart}

\usepackage{iopams}
\usepackage{graphicx}
\usepackage{subfigure}
\usepackage{dcolumn}
\usepackage{bm}
\usepackage{color}
\usepackage{slashed}

\newcommand{\identity}{\mathbb{I}}

\newcommand{\expectation}[1]{\left\langle #1 \right\rangle}

\begin{document}


\title{Quantum correlations and limit cycles in the driven-dissipative Heisenberg lattice}

\author{E. T. Owen}
\address{SUPA, Institute of Photonics and Quantum Sciences, Heriot-Watt University Edinburgh EH14 4AS, United Kingdom}
\address{Joint Quantum Center (JQC) Durham-Newcastle, Department of Physics, Durham University, South Road, Durham DH1 3LE, United Kingdom}
\ead{edmund.t.owen@durham.ac.uk}

\author{J. Jin}
\address{School of Physics, Dalian University of Technology, 116024 Dalian, China}

\author{D. Rossini}
\address{Dipartimento di Fisica, Universit\`a di Pisa and INFN, Largo Pontecorvo 3, I-56127 Pisa, Italy}

\author{R. Fazio}
\address{Abdus Salam ICTP, Strada Costiera 11, I-34151 Trieste, Italy}
\address{NEST, Scuola Normale Superiore \& Instituto Nanoscienze-CNR, I-56126, PISA, Italy}

\author{M. J. Hartmann}
\address{Institute of Photonics and Quantum Sciences, Heriot-Watt University Edinburgh EH14 4AS, United Kingdom}

\date{\today}

\begin{abstract}

Driven-dissipative quantum many-body systems have attracted increasing interest in recent years as they lead to novel classes of quantum many-body phenomena. In particular, mean-field calculations predict limit cycle phases, slow oscillations instead of stationary states, in the long-time limit for a number of driven-dissipative quantum many-body systems. Using a cluster mean-field and a self-consistent Mori projector approach, we explore the persistence of such limit cycles as short range quantum correlations are taken into account in a driven-dissipative Heisenberg model.

\end{abstract}

\maketitle

\section{Introduction}

Understanding the phases of quantum many-body systems is one of the central goals of modern physics.  Phases of matter emerging from cooperative behaviour in equilibrium systems have proven to be of fundamental and technological importance, with notable examples including superconductors~\cite{Micnas:1990} and topological materials~\cite{Qi:2011}.  Recent advances have provided the opportunity to extend this field into the exploration of the phase diagrams of non-equilibrium quantum systems where excitations which dissipate from the system are replenished using an external driving field~\cite{Hartmann:2016, Noh:2017, Sieberer:2016}.  Experimental platforms, such as cavity arrays, superconducting circuits and polariton waveguides, have introduced a new class of systems where the interplay between coherent driving and incoherent dissipation has led to the discovery of novel phenomena.  Bistability~\cite{LeBoite:2013, Fitzpatrick:2017} and crystallisation~\cite{Hartmann:2010} in the driven-dissipative non-linear resonator arrays and synchronised switching in an array of coupled Josephson junctions~\cite{Leib:2014} provide a couple of examples where non-equilibrium phenomena are essential for the understanding of quantum photonic systems.

An intriguing possibility of a non-equilibrium phase that appears in driven-dissipative systems are limit cycles, whereby the system enters a periodic trajectory which breaks the time-translation symmetry of the master equation.  Mean-field studies suggest that limit cycles could exist in a range of non-equilibrium systems, including optomechanical arrays~\cite{Ludwig:2013}, anisotropic Heisenberg lattices~\cite{Lee:2013, Chan:2015}, Rydberg lattices~\cite{Lee:2011, Qian:2012} and Bose-Hubbard lattices with cross-Kerr interactions~\cite{Jin:2013, Jin:2014}.  Experimentally realizing limit cycles would not only be the discovery of a new class of phases in driven dissipative quantum many-body systems but could also have important technological applications, for example in synchronizing quantum many-body devices~\cite{Ludwig:2013, Heinrich:2011}.

Existing predictions of the occurrence of limit cycles are almost exclusively based on Gutzwiller mean-field approaches, which assume a factorized density matrix and ignore quantum fluctuations. It is therefore important to investigate to what extent these limit cycles are affected by quantum fluctuations or correlations~\cite{Navarrete-Benlloch:2017}, which often play a significant role in determining the structure of exotic materials~\cite{Hartmann:2010, Ferrier-Barbut:2016}.  Recently, Chan {\it et al.} have shown that limit cycles persist in the anisotropic Heisenberg lattice even when Gaussian fluctuations are taken into account~\cite{Chan:2015}.  

In this work, we present simulations of the driven-dissipative anisotropic Heisenberg lattice using the self-consistent Mori projector~\cite{Degenfeld-Schonburg:2014} and cluster mean-field methods~\cite{Jin:2016} in order to explore the role of short-range fluctuations beyond the Gaussian approximation.  Within the limits of our approximation limit cycles in the Heisenberg model disappear for reduced dimensionality and increasing cluster size respectively.  Both methods implicitly include fluctuations beyond the Gaussian approximation~\cite{Degenfeld-Schonburg:2014, delValle:2013, Degenfeld-Schonburg:2015, Degenfeld-Schonburg:2016} demonstrating that higher-order correlation functions have an important influence on the existence of limit cycles.  The distinct approaches of these two numerical methods in simulating local quantum correlations complement each other and both show a disappearance of the limit cycle phase of the driven-dissipative anisotropic Heisenberg model at low coordination numbers.

\section{Model}

We investigate the long-time behaviour of the driven-dissipative anisotropic Heisenberg lattice, the phase diagram of which was first studied in Ref.~\cite{Lee:2013}.  The system consists of a regular $d$-dimensional lattice of two-level sites with an energy splitting $\omega_0$ which are coherently driven by an external drive field of strength $\Omega$ and frequency $\omega_D$.  We consider such frequency to be resonant with the energy splitting of the two level system, that is $\omega_D = \omega_0$, make the rotating-wave approximation and move to a frame that rotates at frequency $\omega_0$, such that the uncoupled Hamiltonian is given by
\begin{equation}
    H_0 = \frac{\Omega}{2} \sum_i \sigma_i^x.
\end{equation}
where $\sigma^\alpha_i = \{\sigma^x_i, \sigma^y_i, \sigma^z_i\}$ are the Pauli matrices acting on site $i$ and we have set $\hbar = 1$.  Each site has $z = 2 d$ nearest neighbours which are coupled by an anisotropic Heisenberg term such that the Hamiltonian of the full system is given by
\begin{equation}
    H = H_0 + \sum_{\langle i, j \rangle} \sum_\alpha \frac{J_\alpha}{z} \sigma_i^\alpha \sigma_j^\alpha
\end{equation}
where $z$ is the coordination of the lattice and $\langle i, j \rangle$ denotes nearest-neighbour interactions.  The factor of $z^{-1}$ in the coupling term is required in order to ensure that the energy of the system is extensive.  Additionally, individual sites can spontaneously decay from the excited state to the ground state at a rate $\gamma$.  This gives rise to a Markovian dynamics that is ruled by the following master equation in the Lindblad form
\begin{equation}
    \label{eq:master_equation}
    \dot{R} (t) = -i [H, R(t)] + \frac{\gamma}{2} \sum_i (2 \sigma^-_i R(t) \sigma^+_i - \{\sigma^+_i \sigma_i^-, R(t)\} )
\end{equation}
where $R(t)$ is the density matrix of the whole system and $\sigma^\pm_i$ is the annihilation/creation operator for site $i$.  Avenues to realising this model in experiments with Rydberg atoms and trapped ions have been discussed in Ref.~\cite{Lee:2013}.

\section{Mean-Field Phase Diagram}

Obtaining a solution for Eq.~(\ref{eq:master_equation}) is impractical under almost all circumstances and approximations must be made in order to determine the phase diagram of the model.  The simplest approximation is to ignore quantum correlations between the individual two-level systems and treat interactions as though each site is coupled to the mean-field generated by its nearest neighbours.  Linear stability analysis of the fixed points of the resultant mean-field master equation was performed by Chan {\it et al.}~\cite{Chan:2015} and, for completeness, we summarize some of their results here.

\begin{figure}[t]
    \centering
    \includegraphics[width=0.8\columnwidth]{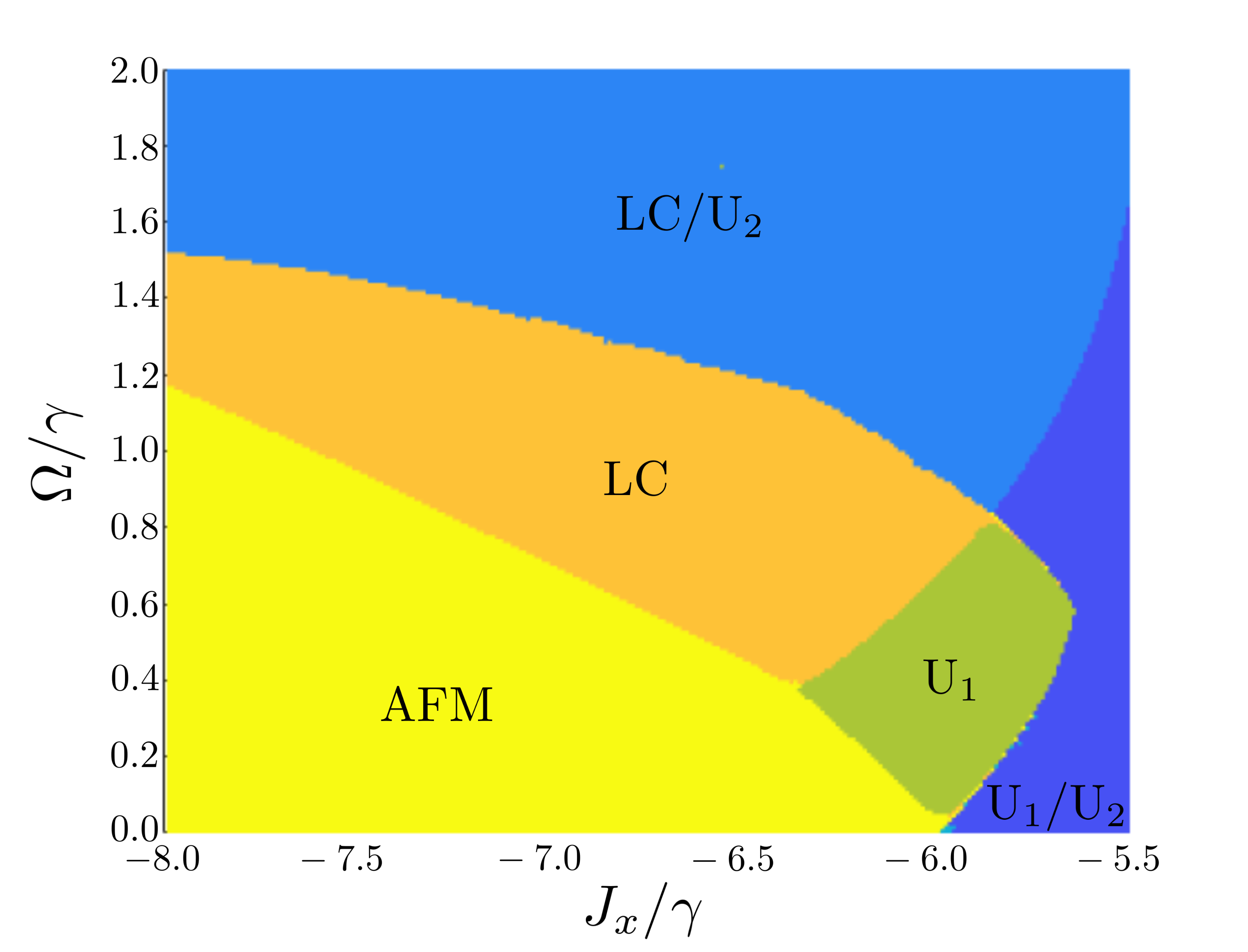}
    \caption{(Color online) Mean-field phase diagram for $\{J_y, J_z\} = \{6.0, 2.0\}$.  The long-time behaviour of the system is separable into antiferromagnetic (AFM), various uniform ($\mathrm{U}_1$ and $\mathrm{U}_2$) and limit cycle (LC) phases.  Bistable regimes where both phases can be reached depending on the system's initial state are also present.
    }
    \label{fig:MF_PhaseDiagram}
\end{figure}

The mean-field phase diagram indicates that an antiferromagnetic phase can be realised in this system.  Therefore, in order to allow the density matrix to break the translational symmetry of the model, we divide the sites into an $A$ and $B$ sublattice where all sites on the $A$ sublattice interact only with the $B$ sublattice and vice versa.  In the mean-field approximation, the equations of motion for the reduced density matrices of the $A$ and $B$ sublattice can then be simplified by replacing the anisotropic Heisenberg coupling with an effective self-consistent classical field generated by the site's $z$ nearest neighbours
\begin{equation}
    \sum_{\langle i, j \rangle} \sum_\alpha \frac{J_\alpha}{z} \sigma_i^\alpha \sigma_j^\alpha \to \sum_\alpha J_\alpha \sigma^\alpha_A \expectation{\sigma^\alpha_B}
\end{equation}
thus excluding quantum correlations.  In Fig.~\ref{fig:MF_PhaseDiagram}, we reproduce a part of the mean-field phase diagram.  Within this approximation, the model supports uniform and antiferromagnetic phases which would also be expected in an equilibrium system.  However, the inclusion of driving and dissipation allows for regions of the phase diagram where the system enters a limit cycle~\cite{Lee:2013, Chan:2015}, see region LC in Fig.~\ref{fig:MF_PhaseDiagram}. 

In the limit cycle phase, the sublattice symmetry of the system is broken and the local system observables of the two sublattices oscillate periodically with a relative phase $\pi$.  The limit cycles exist even though the mean-field Liouvillian is time-independent.  Note that the breaking of the sublattice symmetry is due to the instability of the mean-field solution and is not an artifact of assuming a bipartite lattice~\cite{Lee:2013, Jin:2016}.  In this paper, we explore how the limit cycles predicted by these mean-field calculations are affected by quantum fluctuations and correlations.

\section{Methods}

We explore the existence of limit cycle phases for the model specified in Eq.~(\ref{eq:master_equation}) via two methods, self-consistent Mori projectors and cluster mean field.  

\subsection{Self-consistent Mori projector method}

\label{subsec:cMoP}

The self-consistent Mori projector approach solves for the reduced density matrices of individual lattice sites by integrating out the correlations to give non-Markovian equations of motion for the reduced density matrices of the system~\cite{Degenfeld-Schonburg:2014}.  For the system described by Eq.~(\ref{eq:master_equation}), we start by partitioning the Liouvillian into local and interaction parts
\begin{equation}
    \dot{R} (t) = \mathcal{L}_0 R(t) + \mathcal{L}_I R(t)
\end{equation}
where
\begin{eqnarray}
    \mathcal{L}_0 R (t) & = \sum_j - \frac{i \Omega}{2} [\sigma^x_j, R(t)] + \frac{\gamma}{2} \sum_j (2 \sigma^-_j R(t) \sigma^+_j - \{\sigma^+_j \sigma_j^-, R(t)\} ) \\
    & \equiv \sum_j \mathcal{L}_j R(t) \\
    \mathcal{L}_I R(t) & = \sum_{\langle i, j \rangle} \sum_\alpha \frac{J_\alpha}{z} [\sigma_i^\alpha \sigma_j^\alpha, R(t)].
\end{eqnarray}
The equation of motion for the reduced density matrix $\rho_n$ of site $n$, derived in Ref.~\cite{Degenfeld-Schonburg:2014}, can be expressed as a Dyson series in $\mathcal{L}_I$ and is given by
\begin{eqnarray}
    \dot{\rho}_n (t) &= \mathcal{L}_n \rho_n (t) + \sum_{m = 1}^z \sum_\alpha \frac{J_\alpha}{z} \Tr_{\slashed{n}} \left\{ \left[ \sigma_n^\alpha \sigma_m^\alpha , \bigotimes_k \rho_k (t) \right] \right\} \nonumber \\
    & \qquad + \int_{t_0}^t d t' \Tr_{\slashed{n}} \left\{ \mathcal{L}_I e^{\mathcal{L}_n (t - t')} \mathcal{C}_{t'} \mathcal{L}_I \bigotimes_k \rho_k (t') \right\} \nonumber \\
    \label{eq:cMoP}
    & \qquad + \sum_{m = 3}^\infty \mathcal{Y}^n_m (t)
\end{eqnarray}
where $\Tr_{\slashed{n}} \{ \cdot \}$ denotes the trace over all degrees of freedom except for those of site $n$,  $P_t^n ( \cdot ) = \Tr_{\slashed{n}} \{ \cdot \} \otimes \bigotimes_{m \neq n} \rho_m (t)$ are the time-dependent Mori projectors and $\mathcal{C}_t = \identity - \sum_n P_t^n$.  The first term in Eq.~(\ref{eq:cMoP}) describes the free evolution of the $n$th site whilst the second term is the interaction of the site with the mean field.  The third term is referred to as the Born term and is the first-order quantum correction to the mean-field prediction for the dynamics of $\rho_n (t)$.  The final term is the sum over the remaining terms of the self-consistent Mori projector Dyson series.  Its explicit form can be found in App. C of Ref.~\cite{Degenfeld-Schonburg:2014}.  In order to make simulation of Eq.~(\ref{eq:cMoP}) tractable, we make a truncation of the Dyson series expansion by setting $\mathcal{Y}_m^n (t) = 0$ for $m \geq 3$, corresponding to a Born approximation.  Note that even with this truncation, Eq.~(\ref{eq:cMoP}) was found to give more accurate results than standard perturbation theory to second order, see Fig 3d in~\cite{Degenfeld-Schonburg:2014} where a second order expansion in correlators~\cite{delValle:2013, Li:2014} is compared to self-consistent Mori projector results in the Born approximation. 

With the truncation of Eq.~(\ref{eq:cMoP}) at second order in $\mathcal{L}_I$, we partition the sites onto $A$ and $B$ sublattices such that
\begin{eqnarray}
    \dot{\rho}_A (t) = & \mathcal{L}_A \rho_A(t) - i \sum_\alpha J_\alpha \Tr \{\sigma^\alpha \rho_B(t) \} \left[\sigma^\alpha, \rho_A (t)\right] \nonumber \\
    & \quad - \sum_{\alpha \beta} \frac{J_\alpha J_\beta}{z} \Bigg[ \sigma^\alpha,  \int_0^t d t' \Big( d_{\alpha \beta} (t, t') e^{\mathcal{L}_A (t - t')} \delta \sigma_A^\beta (t') \rho_A (t') \nonumber \\
    \label{eq:system_cMoP}
    & \qquad \qquad \qquad \quad - s_{\alpha \beta} (t, t') e^{\mathcal{L}_A (t - t')} \rho_A (t') \delta \sigma_A^\beta (t') \Big) \Bigg]
\end{eqnarray}
where
\begin{equation}
    d_{\alpha \beta} (t, t') = \Tr \{ \sigma^\alpha e^{\mathcal{L}_B (t - t')} \delta \sigma_B^\beta (t') \rho_B (t') \}
\end{equation}
\begin{equation}
    s_{\alpha \beta} (t, t') = \Tr \{ \sigma^\alpha e^{\mathcal{L}_B (t - t')} \rho_B (t') \delta \sigma_B^\beta (t') \}
\end{equation}
\begin{equation}
    \delta \sigma_A^\beta (t) = \sigma^\beta - \Tr \{ \sigma^\beta \rho_A (t) \} \identity_2
\end{equation}
\begin{equation}
    \delta \sigma_B^\beta (t) = \sigma^\beta - \Tr \{ \sigma^\beta \rho_B (t) \} \identity_2
\end{equation}
and similarly for $\rho_B (t)$.  For simplicity, we have omitted the site index for the Pauli operators, which operate on the relevant sublattice.   The long-time behaviour of the system can then be calculated by time-integrating the equations of motion over a sufficiently long time such that the transient behaviour has disappeared.

\subsection{Cluster mean-field theory}

Approximate solutions to the master equation become more accurate when reduced density matrices for clusters consisting of multiple lattice sites are considered.~\cite{Degenfeld-Schonburg:2014, Jin:2016}.  Quantum correlations within these clusters are then calculated exactly and inaccuracies are limited to interactions between clusters.  The exponential scaling of the dimension of the Hilbert space with increasing coordination number unfortunately makes it impossible to consider clusters for  lattices with large $z$ but, for low coordination number, simulating the reduced density matrix of a cluster becomes more manageable as the size of the cluster's Hilbert space is reduced.  However, when using the self-consistent Mori projector method for this model, evaluating the Born term in Eq.~(\ref{eq:cMoP}) for a cluster is computationally difficult due to the large number of terms in the interaction Liouvillian.  Nevertheless, even in a mean-field calculation, all correlations are taken into account for the internal quantum dynamics of the cluster which interacts with the mean-field exerted by its neighbouring clusters.

In the cluster mean-field approximation, the density matrix of the lattice is divided into a product state of contiguous clusters of sites $\mathcal{C}$ which are identical due to the translational symmetry of the lattice
\begin{equation}
    R (t) \approx \bigotimes_i \rho_{\mathcal{C}}.
\end{equation}
This density matrix then evolves according to the decoupled cluster mean-field Liouvillian which can be written as
\begin{equation}
    \mathcal{L}_{\mathrm{CMF}} = \mathcal{L}_{\mathcal{C}} + \mathcal{L}_{\mathcal{B(C)}}
\end{equation}
where $\mathcal{L}_{\mathcal{C}} = \sum_{j \in \mathcal{C}} \mathcal{L}_j$ describes the evolution of the isolated cluster and the interaction with the mean-field of the neighbouring clusters is described by the non-linear Liouvillian $\mathcal{L}_{\mathcal{B(C)}}$ which acts only on sites at the boundary of the cluster.  For the driven-dissipative Heisenberg model, the boundary Liouvillian is given by
\begin{equation}
    \mathcal{L}_{\mathcal{B(C)}} = \sum_{j \in \mathcal{B(C)}} -i \left[ \mathbf{B}_j^{\mathrm{eff}} (t) \cdot {\bm \sigma}_j, \rho_{\mathcal{C}} \right]
\end{equation}
where $\mathbf{B}_j^{\mathrm{eff}} (t)$ is related to the average polarization of the sites adjacent to the boundary $\mathcal{B (C)}$ at time $t$.  Clusters containing an odd number of sites break the bipartite symmetry of the lattice and a pair of complementary coupled clusters must be used.  For example, in order to perform calculations using a cluster size of $3 \times 3$, the lattice can be divided into two subclusters $\mathcal{C}_A$ and $\mathcal{C}_B$ where the four sites in the corners and the central site of $\mathcal{C}_A$ are assigned to the $A$ sublattice and $\mathcal{C}_B$ is the complement of $\mathcal{C}_A$.

\section{Results}

\begin{figure}[t]
    \centering
    \includegraphics[width=0.8\columnwidth]{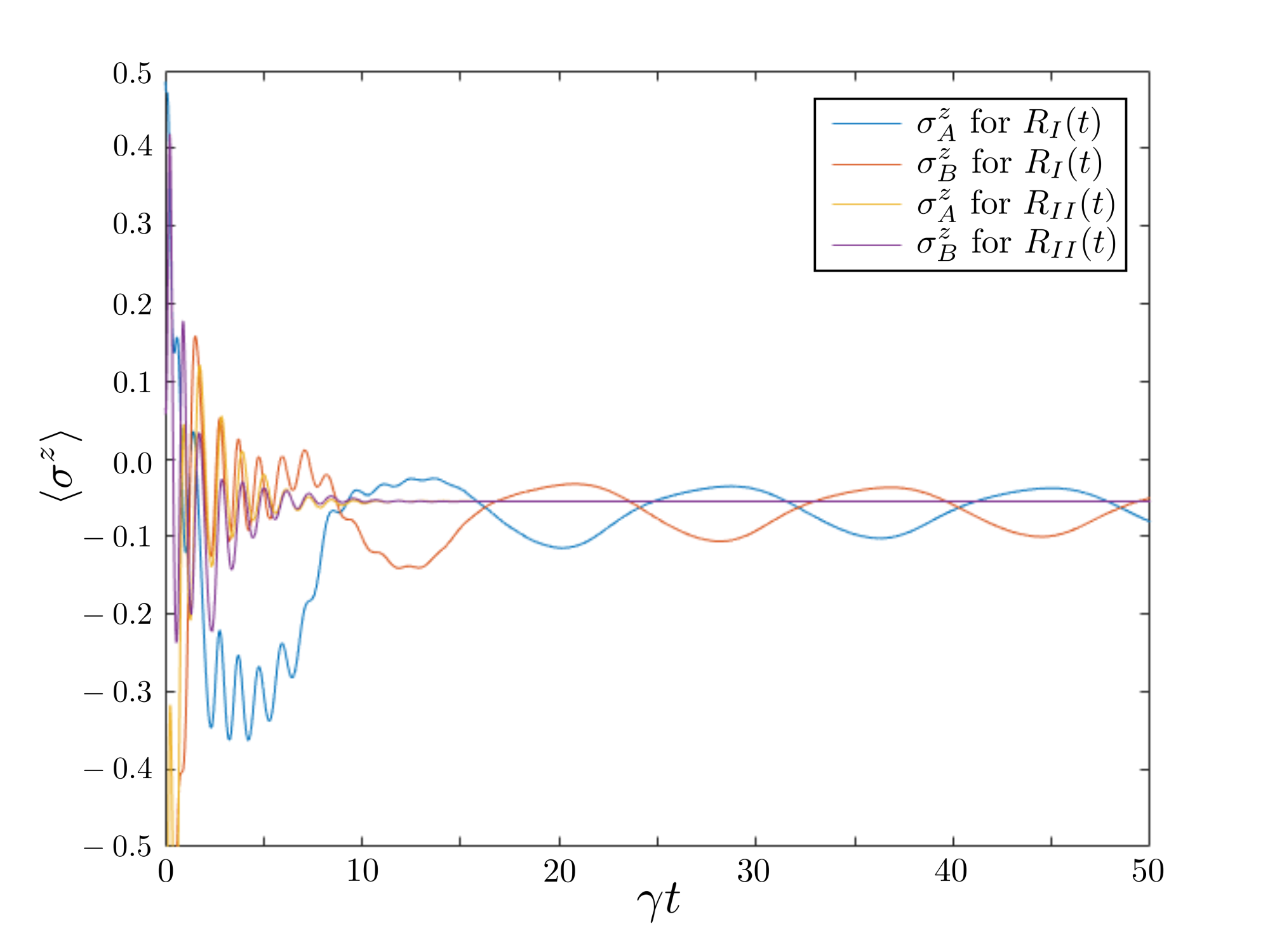}
    \caption{(Color online) An example of the dynamics of two sample initial product states for $z = 150$ for $\{J_x, J_y, J_z, \Omega\} = \{-7.0, 6.0, 2.0, 1.0\}$.  The system was initiated in the random product state where for the initial state $R_{\mathrm{I}}(0)$ which relaxes to a stationary steady state, $\mathbf{n}_A = \{0.0911, -0.5318, -0.7725\}$ and $\mathbf{n}_B = \{-0.0007, -0.6958, 0.0654\}$ and for the initial state $R_{\mathrm{II}}(0)$ which enters the limit cycle phase, $\mathbf{n}_A = \{0.2576, 0.1597, 0.1999\}$ and $\mathbf{n}_B = \{-0.4684, -0.4306, -0.4928\}$.
    }
    \label{fig:LC_example}
\end{figure}

We apply the methods described in the previous section to find the long-time behaviour of the driven-dissipative Heisenberg lattice.

At $t = 0$, the system is prepared in a product state, that we parametrise as
\begin{equation}
    \label{eq:initial_state}
    R(0) = \bigotimes_{i \in A} \frac{1}{2} \left(\identity_2 + \sum_\alpha n^\alpha_A \sigma^\alpha \right)_i \ \bigotimes_{j \in B} \frac{1}{2} \left(\identity_2 + \sum_\alpha n^\alpha_B \sigma^\alpha \right)_j
\end{equation}
where $n^\alpha_A$ and $n^\alpha_B$ are the components of two vectors within the unit sphere.  For this initial state, we time-integrate the equations of motion of the respective reduced density matrices using both the self-consistent Mori projector method and cluster mean-field theory until $t \gg \gamma^{-1}$, where the transient behaviour due to the product state initialisation has decayed and the system has either entered into a limit cycle or has reached a time-independent steady state.  We performed the simulations for two sets of parameters: $\{J_x, J_y, J_z, \Omega\} = \{-7.0, 6.0, 2.0, 1.0\}$ and $\{J_x, J_y, J_z, \Omega\} = \{-6.4, 3.0, 6.0, 2.25\}$ with $\gamma = 1$ which both correspond to points in the parameter space where mean-field calculations predict the steady state to be composed of limit cycles.  

\subsection{Self-Consistent Mori Projector Method}

\begin{figure}[t]
    \centering
    \includegraphics[width=0.8\columnwidth]{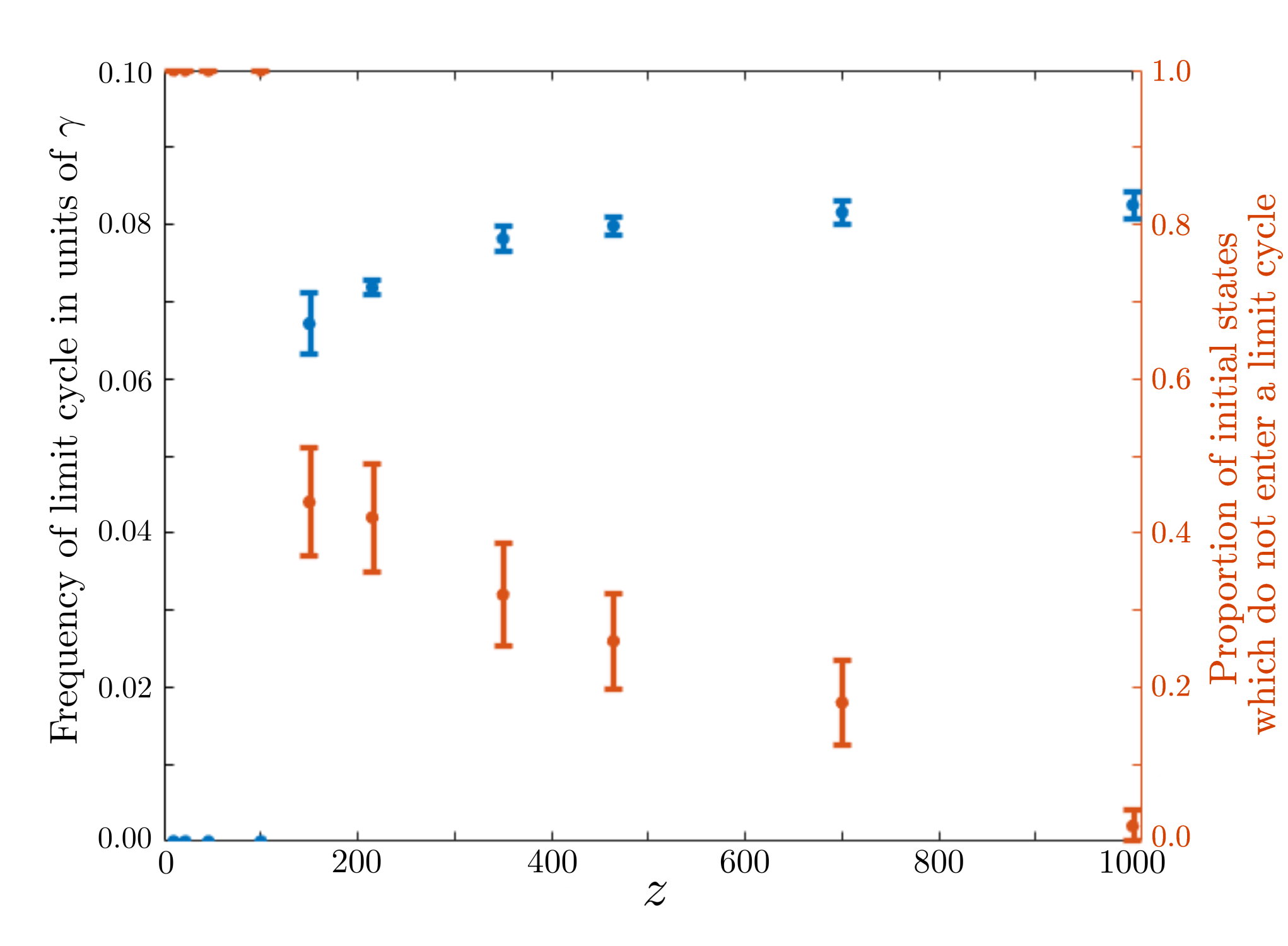}
    \caption{(Color online) Frequency and proportion of initial states entering the limit cycle phase as a function of the coordination number for $\{J_x, J_y, J_z, \Omega\} = \{-7.0, 6.0, 2.0, 1.0\}$.  Below a critical dimension $z^* \approx 100$, the limit cycle phase disappears and all initial states relax to a paramagnetic stationary state.  This transition is accompanied by a shift in the frequency of the limit cycle.  The error in the frequency is due to computational limitations which restricted the period of time over which the density matrix could be evolved whilst the uncertainty in the proportion of initial states not entering a limit cycle is taken from the standard error of a Bernoulli process.
    }
    \label{fig:LC_freq}
\end{figure}

\begin{figure}[t]
    \centering
    \includegraphics[width=0.8\columnwidth]{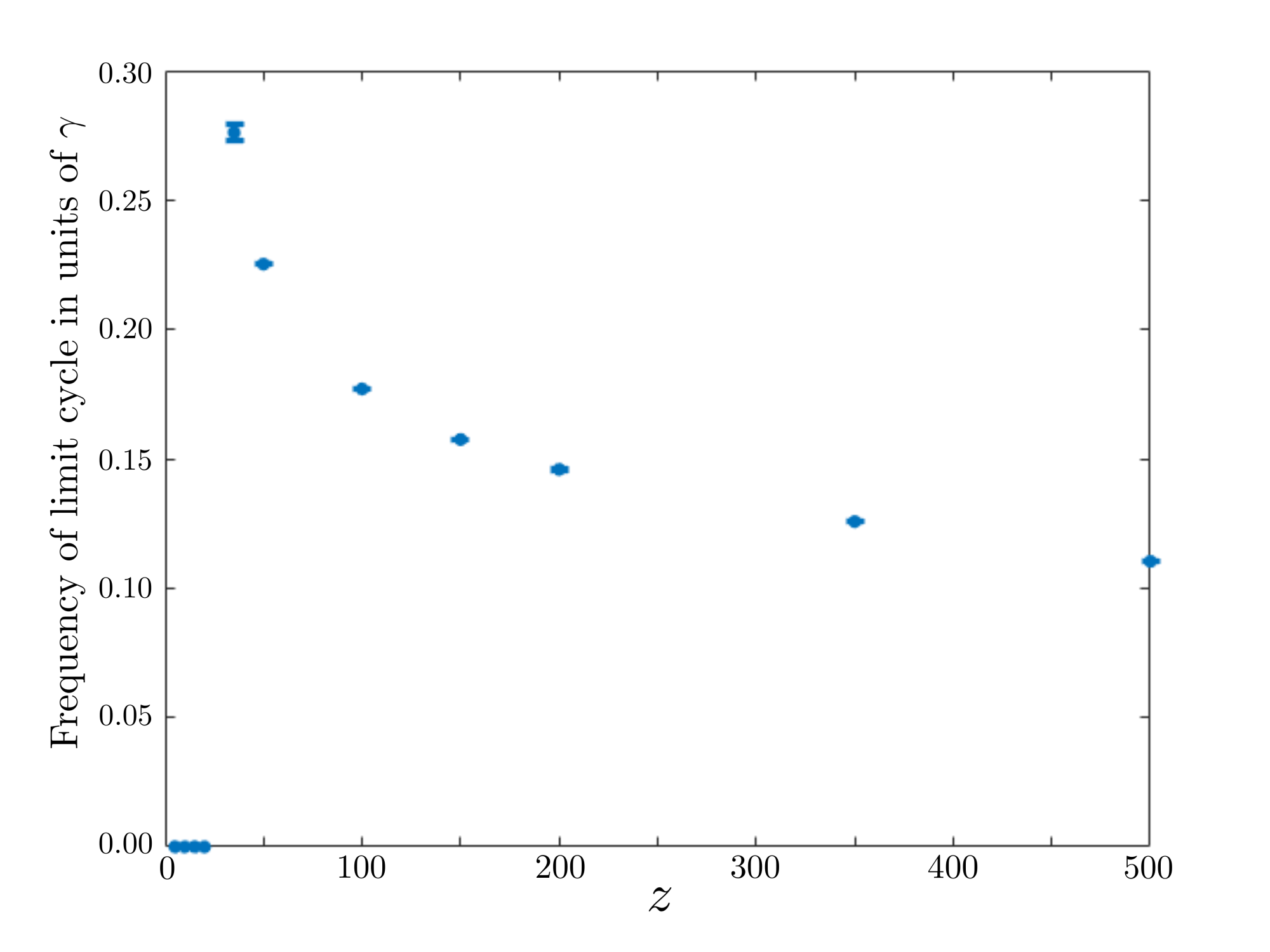}
    \caption{(Color online) Frequency of initial states entering the limit cycle phase as a function of the coordination number for $\{J_x, J_y, J_z, \Omega\} = \{-6.4, 3.0, 6.0, 2.25\}$.  The critical dimension is lower than for the parameters used in Fig.~\ref{fig:LC_freq} with $z^* \approx 50$, but below this value, all initial states still relax to a paramagnetic stationary state.  For $z > z^*$, all initial states enter into a limit cycle.
    }
    \label{fig:LC_freq_Robust}
\end{figure}

We start by presenting the results calculated using the self-consistent Mori projector method.  The system exhibits two distinct long-time behaviours which we show in Fig.~\ref{fig:LC_example}, where we characterize the limit cycle using the average polarization of the sublattices $\Tr \{\sigma^z \rho_A\}$ and $\Tr \{\sigma^z \rho_B\}$.  For some initial states, after a transient behaviour, the sublattice symmetry is restored and the system relaxes to a ferromagnetic stationary state.  However, for the other initial states, the system enters a limit cycle where the sublattice symmetry is broken and the system oscillates anharmonically.

Whether a random initial state will enter the limit cycle phase depends strongly both on the system parameters and the coordination number of the lattice.  Figs.~\ref{fig:LC_freq} and~\ref{fig:LC_freq_Robust} show the proportion of states which do not enter the limit cycle phase as a function of the coordination number $z$ from a sample of 50 initial states with randomised $n_A^\alpha$ and $n_B^\alpha$.  In the limit $z \to \infty$, the equations of motion become equivalent to the mean-field approximation as the prefactor of the Born term in Eq.~(\ref{eq:system_cMoP}) vanishes so all initial states enter into limit cycles.  Whilst this limit is unlikely to be experimentally practical, it does allow us to connect the mean-field result to more accurate investigations as the coordination number of the system controls the magnitude of the Born term and all higher order terms $\mathcal{Y}^n_m (t)$ with $m \ge 3$ in Eq.~(\ref{eq:cMoP}).  As the coordination number of the lattice decreases, quantum correlations included within the Mori projector expansion become relevant and certain initial states will evolve towards a stationary ferromagnetic state.  For the above parameters, we find that below a critical coordination number $z^*$, the limit cycles are absent and there is no evidence of long-time oscillatory behaviour for all initial states.  The critical coordination number differs depending on the system parameters but $z^* > 10$ in both cases. In contrast to $\{J_x, J_y, J_z, \Omega\} = \{-7.0, 6.0, 2.0, 1.0\}$, for $\{J_x, J_y, J_z, \Omega\} = \{-6.4, 3.0, 6.0, 2.25\}$ all initial states enter limit cycles for $z > z^*$.  

In the mean-field limit, we see that for these parameter sets, every initial state is attracted to a limit cycle.  As the coordination number of the system decreases, a proportion of these initial states instead are attracted to a time-independent steady state.  For $z < z^*$, our numerical results indicate that the size of the attractor basin for the limit cycle phase disappears and all initial states converge on a stationary solution.  This result indicates that it is not possible to enter into a limit cycle phase for experimentally realistic coordination numbers.

Whilst this transition is accompanied by a shift in the frequency of the limit cycle, it is not possible to perform a quantitative analysis of this shift as higher-order terms in Eq.~(\ref{eq:cMoP}) may become relevant.  For the parameters considered here, the truncated terms in the Dyson series given in Eq.~(\ref{eq:cMoP}) scale as $\mathcal{Y}_m^n (t) \sim (J_\alpha / \gamma)^m$ and therefore convergence is not guaranteed as $|J_\alpha / \gamma| \geq 1$ for $\alpha = x, y$ and $z$.  Hence, the difference in behaviour of the limit cycle frequency between the two parameter sets presented here for coordination numbers close to the critical value $z^*$ can not necessarily be expected to be a quantitatively reliable prediction.  While we cannot exclude the possibility that higher order terms would restore the limit cycle behavior, we do not expect that expanding Eq.~(\ref{eq:cMoP}) to first order in the couplings -- which is the mean-field approximation -- will be more accurate than the expansion to second order that we consider here.

\subsection{Cluster mean-field theory}

\label{subsec:ClusterMF}

\begin{figure}[t]
    \centering
    \includegraphics[width=0.8\columnwidth]{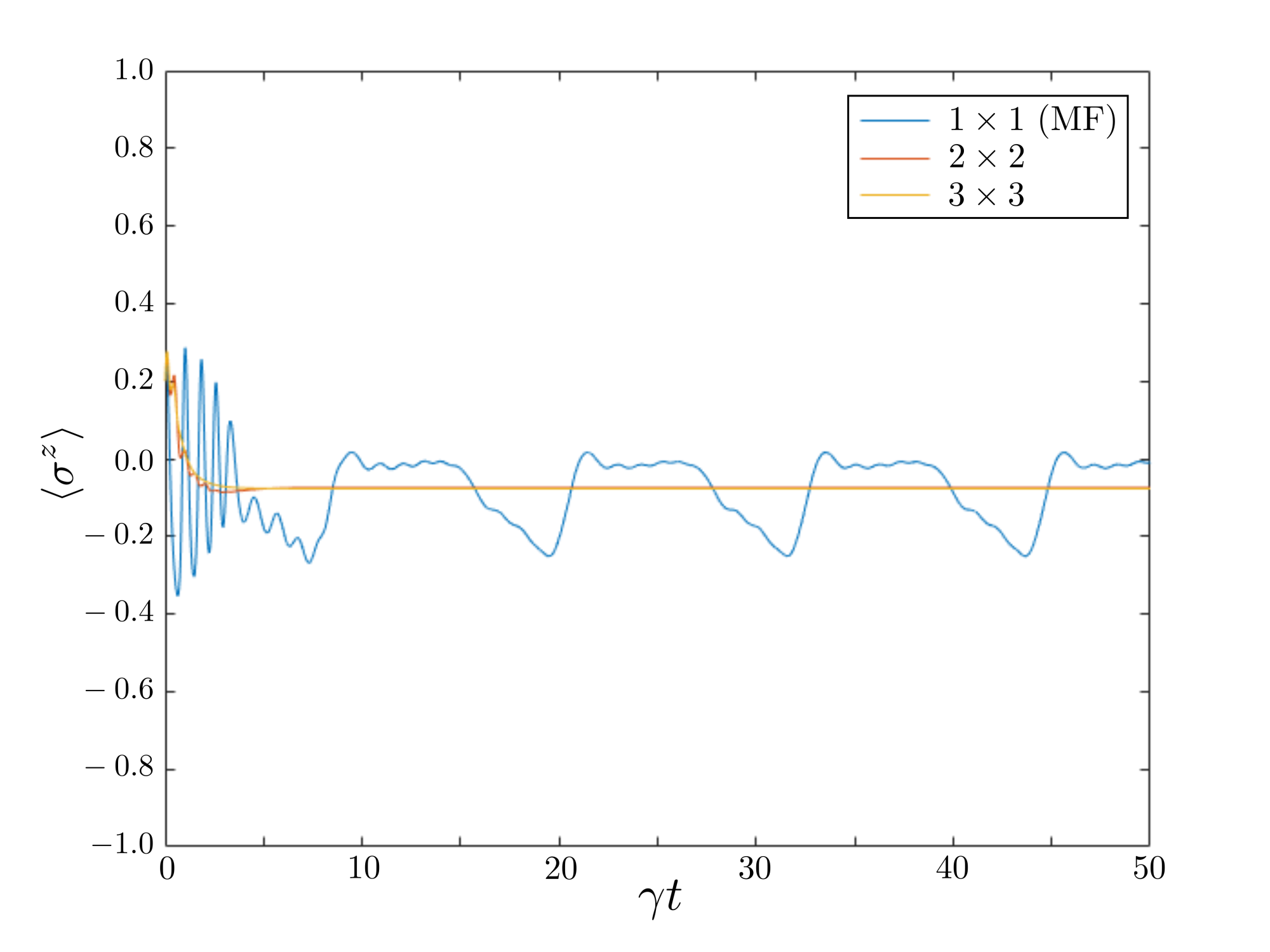}
    \caption{(Color online)  Time-evolution of the polarisation of the two-dimensional driven-dissipative Heisenberg model using cluster mean-field theory for $\{J_x, J_y, J_z, \Omega\} = \{-7.0, 6.0, 2.0, 1.0\}$ for which the system was initiated in the product state $R_{\mathrm{II}}(0)$ as defined in Fig.~\ref{fig:LC_example}.  Simulations using $2 \times 2$ and $3 \times 3$ clusters both converge on a time-independent steady state in contrast to the mean-field prediction.
    }
    \label{fig:CMF_2D}
\end{figure}

\begin{figure}[tph]
    \centering
    \includegraphics[width=0.8\columnwidth]{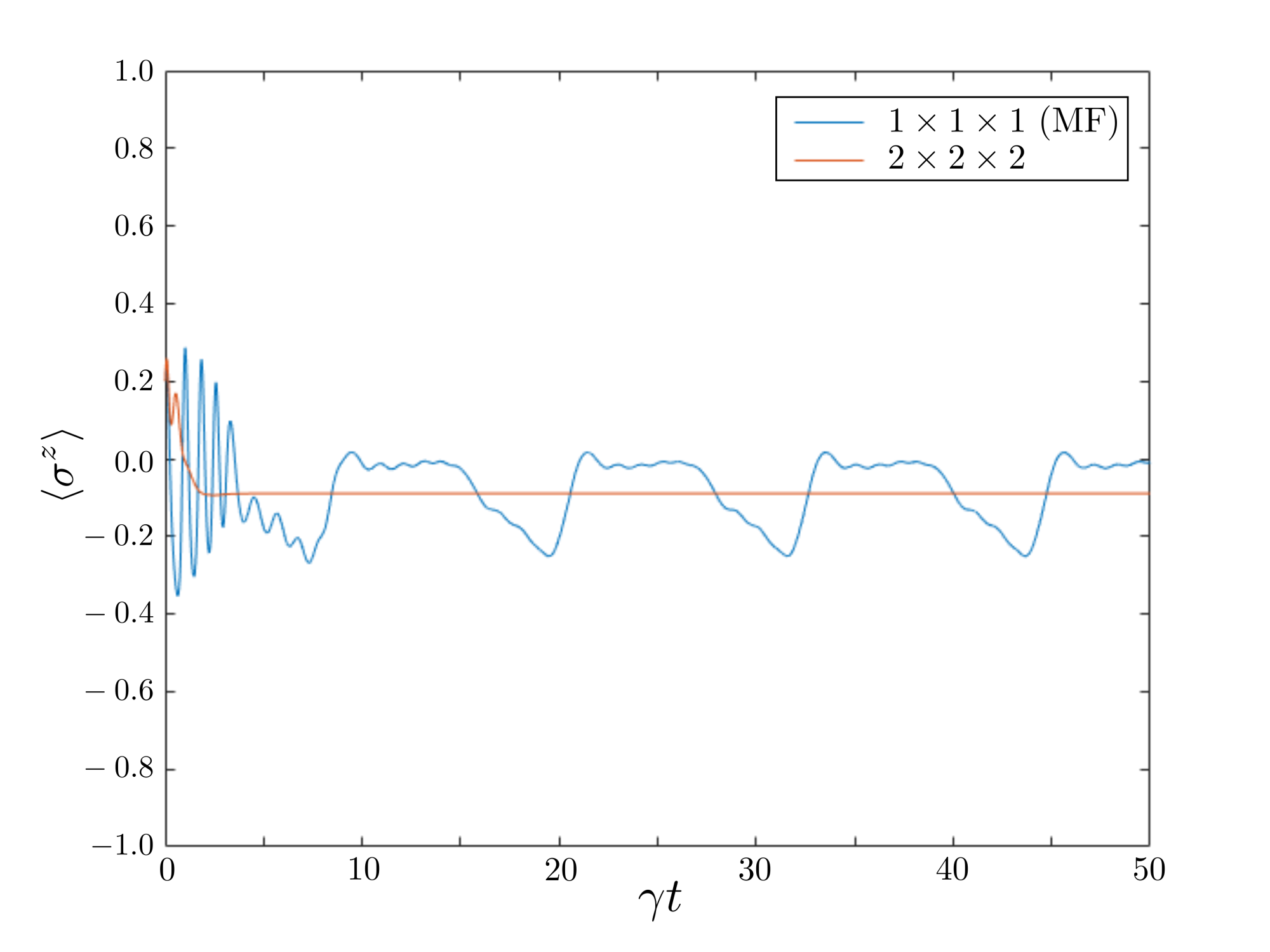}
    \caption{(Color online)  Time-evolution of the polarisation of the three-dimensional driven-dissipative Heisenberg model using cluster mean-field theory for $\{J_x, J_y, J_z, \Omega\} = \{-7.0, 6.0, 2.0, 1.0\}$ for which the system was initiated in the product state $R_{\mathrm{II}}(0)$ as defined in Fig.~\ref{fig:LC_example}.  Similarly to the two-dimensional model, results from larger clusters do not exhibit limit cycles.
    }
    \label{fig:CMF_3D}
\end{figure}

Next we investigate the existence of limit cycle phases in the driven-dissipative anisotropic Heisenberg model via cluster mean-field theory.  Fig.~\ref{fig:CMF_2D} shows the average polarisation of the $A$ and $B$ sublattices for a two-dimensional lattice calculated using cluster mean-field simulations of $2 \times 2$ and $3 \times 3$ clusters.  The wave function was initiated in the product state given by Eq.~(\ref{eq:initial_state}) for $\{J_x, J_y, J_z\} = \{-7.0, 6.0, 2.0, 1.0\}$. Fig.~\ref{fig:CMF_2D} shows that the limit cycles are not observed in the cluster mean-field simulations.  Such conclusions extend to cluster mean-field simulations for a three dimensional lattice.  In Fig.~\ref{fig:CMF_3D}, we show the average polarisation for a $2 \times 2 \times 2$ cluster.  Once again, the system relaxes into a stationary steady state after an initial transient and limit cycles are not observed. 

\begin{figure}[bph]
    \centering
    \includegraphics[width=0.8\columnwidth]{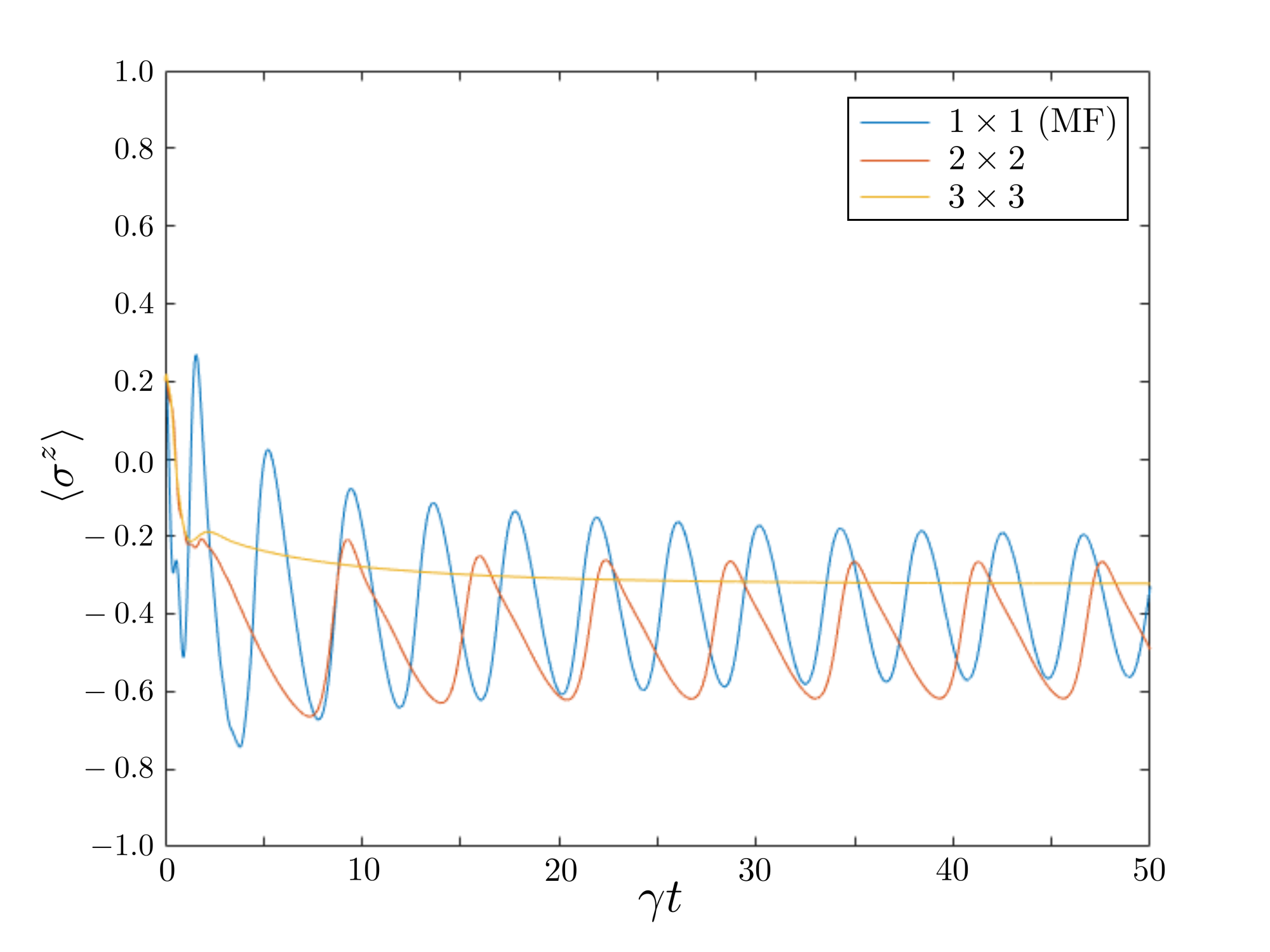}
    \caption{(Color online)  Time-evolution of the polarisation of the two-dimensional driven-dissipative Heisenberg model using cluster mean-field theory for $\{J_x, J_y, J_z, \Omega\} = \{-6.4, 3.0, 6.0, 2.25\}$ for which the system was initiated in the product state $R_{\mathrm{II}}(0)$ as defined in Fig.~\ref{fig:LC_example}.  The limit cycles are more robust than in Fig.~\ref{fig:CMF_2D}, as they persist for $2 \times 2$ clusters, but the system relaxes to a stationary steady state for $3 \times 3$ clusters.
    }
    \label{fig:CMF_2D_Robust}
\end{figure}

For $\{J_x, J_y, J_z, \Omega\} = \{-6.4, 3.0, 6.0, 2.25\}$, cluster mean-field theory shows that the limit cycle phase is more robust as the system still exhibits periodic behaviour for a $2 \times 2$ cluster (see Fig.~\ref{fig:CMF_2D_Robust}).  However, for a $3 \times 3$ cluster, once again the system relaxes to a stationary steady state.

\section{Conclusions}

We have used the self-consistent Mori projector and cluster mean-field methods to simulate the evolution of a driven-dissipative anisotropic Heisenberg model within a regime where mean-field results predict that this system should exhibit limit cycles.  The approximations required for these two methods are limited to calculating local properties of the system but have the advantage of including higher-order quantum correlation effects.  For a high coordination number, the self-consistent Mori projector method agrees with the mean-field prediction and limit cycles are observed for any initial state.  However, as the coordination number of the lattice is reduced, a proportion of the initial states no longer enter the limit cycle phase and relax towards a ferromagnetic steady state. Below a critical coordination number, for which $z^* \gtrsim 10$ for the parameter sets which we have studied here, the limit cycle phase disappears and all initial states relax towards the same steady state.  Recently limit cycles in many-body systems were associated with the appearance of time-crystalline behavior~\cite{Iemini:2017}. Our work seems to indicate that these effects may appear only in long-range systems or for high dimensions.

\section{Acknowledgements}

ETO and MJH were supported by the EPSRC via EP/N009428/1.  ETO acknowledges funding from the SFC PECRE scheme coordinated by SUPA. JJ acknowledges support from the National Natural Science Foundation of China No. 11605022 and No. 11547119, Natural Science Foundation of Liaoning Province No. 2015020110, and the Xinghai Scholar Cultivation Plan and the Fundamental Research Funds for the Central Universities.  RF acknowledges EU-QUIC and MIUR-QUANTRA. 

\vspace{1.0cm}

\bibliographystyle{unsrt}
\bibliography{citations}

\end{document}